\documentstyle[12pt,equation,epsfig]{article}
\begin{document}


\newcommand{\be}{\begin{equation}}
\newcommand{\ben}{\begin{subequations}}
\newcommand{\een}{\end{subequations}}
\newcommand{\beq}{\begin{eqalignno}}
\newcommand{\eeq}{\end{eqalignno}}
\newcommand{\ee}{\end{equation}}
\newcommand{\epem}{\mbox{$e^+ e^-$}}
\newcommand{\tchi}{\mbox{$\tilde \chi_1^0$}}
\newcommand{\mchi}{\mbox{$m_{\tilde \chi}$}}
\newcommand{\st}{\mbox{$\tilde{t}_1$}}
\newcommand{\tp}{\mbox{$\tilde{t}_1 \tilde{t}_1^*$}}
\newcommand{\tm}{\mbox{$\tilde{t}_M$}}
\newcommand{\mt}{\mbox{$m_{\tilde{t}_1}$}}
\newcommand{\rs}{\mbox{$\sqrt{s}$}}
\renewcommand{\thefootnote}{\fnsymbol{footnote}}


\pagestyle{empty}
\begin{flushright}
IFT--P.018/99 \\
MADPH--99--1105\\
February 1999\\
\end{flushright}
\vspace*{2cm}
\begin{center}
{\Large \bf Strong Interaction Effects in Stop Pair Production at
$e^+ e^-$ Colliders} \\
\vspace*{6mm}
Manuel Drees$^1$ and Oscar J.P. \'Eboli$^{1,2}$\\
$^1${\it Instituto de F\'{\i}sica Te\'orica , Universidade. Estadual
Paulista,\\
Rua Pamplona 145, S\~ao Paulo, SP 01405-900, Brazil} \\
$^2${\it Department of Physics, University of Wisconsin, \\
Madison, WI 53706, USA}
\end{center}

\vspace*{1cm}
\begin{abstract}

    We discuss perturbative and non--perturbative strong interaction effects
    in the pair production of stop squarks (\st) at \epem\ colliders. Events
    with an additional hard gluon allow to detect or exclude \tp\ production
    even in scenarios with very small mass splitting between \st\ and an
    invisible lightest supersymmetric particle (LSP). Such events can also
    help to establish that \st\ transforms as a triplet under $SU(3)_C$. We
    also carefully study non--perturbative \st\ fragmentation, which is
    currently not well understood: not only is the \st\ fragmentation function
    not known very well, but also there are ambiguities in the algorithm
    employed to model fragmentation. We present numerical results both for
    CERN LEP--183 and for a proposed future \epem\ collider operating at
    center--of--mass energy $\rs=500$ GeV.

\end{abstract}

\vspace*{3mm}
\noindent

\clearpage
\pagestyle{plain}
\setcounter{page}{1}


\section*{1) Introduction}

The lighter scalar top (stop) mass eigenstate \st\ is likely to be the
lightest squark \cite{1}. In models where supersymmetry breaking is
communicated to the superpartners of the known SM particles at some energy
well above the weak scale, renormalization group effects reduce the soft mass
of the stops compared to those of other squarks. In addition, after the
spontaneous breakdown of the electroweak $SU(2) \times U(1)_Y$ symmetry,
$SU(2)$ doublet and singlet stops mix, leading to a further reduction of the
mass \mt\ of the lighter eigenstate. In fact, \st\ might even be the lightest
charged sparticle. It is therefore actively being searched for at both $p
\bar{p}$ \cite{2} and \epem\ \cite{3,4} colliders.  In this article we focus
on \tp\ production at existing and future \epem\ colliders.

Unlike the other candidates for the lightest charged superparticle, sleptons
and the lighter chargino mass eigenstate $\tilde{\chi}_1^\pm$,
\st\ has strong interactions. Stop pairs can thus be produced in association
with a hard gluon. Events of this type can help to overcome Standard
Model backgrounds to stop production in scenarios where \st\ decays
into a lightest neutralino \tchi\ and a quark, if \tchi\ is invisible
(stable or long--lived) and $\mt - \mchi$ is small. In such cases the
emission of a hard gluon can increase the total visible energy as well
as the total missing transverse momentum, helping to discriminate 
between \tp\ events and backgrounds from two--photon reactions. Once
\st\ has been discovered, one will also want to study its properties.
Events with an additional hard gluon will then be crucial in establishing
the color charge of \st.

If $\mt < m_b + m_{\tilde{\chi}_1^\pm}, m_t + \mchi$, \st\ is expected to be
quite long--lived \cite{5}. It will then fragment into a ``stop meson'', which
is a spin--1/2 particle, plus a fragmentation jet prior to its decay. This is
a non--perturbative process, and hence not well understood. In fact, the very
concept of fragmentation is somewhat ill--defined for massive particles:
whereas one massless parton can fragment into two collinear massless partons
while conserving energy and momentum, a massive particle cannot fragment into
another particle with equal (or larger) mass and a second, massless particle
with nonvanishing energy. This does not pose much of a problem in the
calculation of single--particle inclusive cross sections, through which
fragmentation functions are usually defined, but it leads to ambiguities in
the implementation of fragmentation in event generator programs. This is of
special importance in scenarios with small \st--\tchi\ mass splitting, where
the fragmentation jets may contribute the bulk of the visible energy.

Since non--perturbative \st\ fragmentation occurs even if no hard gluon is
emitted, we begin by discussing this process in the simple case of $\epem
\rightarrow \tp$. We introduce three different fragmentation schemes, which
differ in the precise definition of the fragmentation (scaling) variable $x$,
and/or in the algorithm used to ensure overall energy--momentum
conservation. These three algorithms lead to quite different stop energy
distributions after fragmentation when used naively with the same
fragmentation function. On the other hand, these differences become quite
small if the parameter of the fragmentation function is adjusted such that the
mean stop ``meson'' energy is the same in all three scenarios.

In Sec.~3 we discuss events where one hard gluon is emitted. We give an
explicit expression for the differential cross section in a form that is
easily implementable in an event generator. We then demonstrate, using a
parton--level simulation, that the possibility of emitting a hard gluon allows
to discover or rule out \st\ even for very small mass splitting to the LSP. To
this end we employ a set of cuts modeled on the OPAL analysis \cite{4}. These
cuts also greatly reduce the sensitivity to the details of \st\ fragmentation.
We also present a second analysis for a future \epem\ collider operating at
center--of--mass energy $\rs=500$ GeV, as an example of the type of analysis
one might perform to determine the color charge of \st. In this case the cross
section after cuts does depend on the details of fragmentation, but the
resulting ambiguity in the prediction is much smaller than the perturbatively
calculable contribution. Finally, Sec.~4 is devoted to a brief summary and
conclusions.


\section*{2) Fragmentation}

We begin our analysis with a discussion of \st\ fragmentation. We will model
it using the Peterson {\it et al.} fragmentation function \cite{6}:
\be \label{e1}
D_{\tilde t}(x) = \frac {1} {N} \frac {1} { x \left( 1 - \frac{1}{x}
- \frac {\epsilon_{\tilde t}}{1-x} \right)^2 } \;\;,
\ee
where the single free parameter $\epsilon_{\tilde t}$ is expected to scale as
the inverse of the square of the stop mass
\be \label{e2}
\epsilon_{\tilde t} = \frac {\epsilon} {m^2_{\tilde t_1}},
\ee
with $\epsilon \sim 0.1$ to $0.5$ GeV$^2$. $D_{\tilde t}$ has to satisfy the
``stop number conservation'' condition $\int_{x_{\rm min}}^1 dx~x D_{\tilde
t}(x) = 1$, where $x_{\rm min}$ is the kinematic minimum of $x$.  This leads
to
\be \label{e3}
N(x_{\rm min}) = \frac {\pi} {4 \sqrt{\epsilon_{\tilde t}}} - \frac{1}{2}
\log \frac {(1 - x_{\rm min})^2} {\epsilon_{\tilde t}} + 1 -
\frac {1} { 2 (1-x_{\rm min}) } + {\cal O}(\sqrt{\epsilon_{\tilde t}}) \;\;.
\ee

Fragmentation functions are usually defined through differential 
single--particle inclusive cross sections. In the simple case of
leading order stop pair production, $\epem \rightarrow \tp$, the stop
energy prior to fragmentation is fixed\footnote{We ignore initial state
radiation in our analysis. It can reduce the total cross section by
$\sim 20$\% \cite{7}, but will not change most kinematical distributions
very much, which are the main emphasis of this work.} to $\rs/2$, so
that
\be \label{e4}
\frac {d \sigma} {d Q} \propto D_{\tilde t}(x) \;\;.
\ee
Unfortunately the choice of $Q$, and consequently of the associated
fragmentation variable $x$, is somewhat ambiguous for massive particles.  It
can be the energy $E$ of the stop ``meson'' \tm\ after fragmentation, in which
case
\be \label{e5}
x = x_E \equiv \frac {E_{\tilde{t}_M}}{E_{\tilde{t}_1}} 
= \frac {2 E_{\tilde{t}_M}} {\sqrt{s}} \;\;;
\ee
or one could choose the absolute value of the 3--momentum $\vec{k}$,
\be \label{e6}
x = x_k \equiv \frac {\left| \vec{k}_{\tilde{t}_M} \right| } 
{\left| \vec{k}_{\tilde{t}_1} \right|} 
= \frac {2 \left| \vec{k}_{\tilde{t}_M} \right|} {\beta \sqrt{s}} \;\;,
\ee
where $\beta = \sqrt{1 - 4 m^2_{\tilde{t}_1} / s }$. The second identities in
eqs.~(\ref{e5}) and (\ref{e6}) hold only for the simple case of $2 \rightarrow
2$ kinematics. Occasionally the fragmentation variable is also defined through
the ``light cone variable'' $(E + |\vec{k}|)/2$.  This will give results
intermediate between the other two choices, so we focus on definitions
(\ref{e5}) and (\ref{e6}) here.

A second ambiguity is related to the fact that a massive (s)quark cannot
fragment into a hadron of equal or greater mass and one (or more) massless
parton(s) without violating energy and/or momentum conservation, if all
particles are on--shell. It is therefore not obvious how to introduce
fragmentation functions, defined through inclusive cross sections as in
eq.~(\ref{e4}), into event generator programs that attempt to describe all
(relevant) properties of events. We have implemented two different classes of
algorithms that ensure energy and momentum conservation at the level of the
global event. For simplicity we always model the ``fragmentation jet'' as a
single massless parton collinear with the parent stop.

The first class of algorithms is based on rescaling the 3--momenta of {\em
all} particles in the final state by some overall factor, after the stops have
been fragmented. The value of this factor is calculated for each event through
iteration, by requiring that all final particles are on--shell and that their
energies in the \epem\ center--of--mass (cms) frame add up to \rs. These
algorithms require that the 3--momenta of all particles after fragmentation,
but before rescaling, add up to zero (in the \epem\ cms frame); overall
3--momentum conservation is then ensured throughout the rescaling
process. This means that the fragmentation step itself always has to fix the
3--momentum of the fragmentation jet, even if the fragmentation variable is
defined through the stop energy as in eq.~(\ref{e5}). In other words, this
fragmentation prescription ``locally'' conserves 3--momentum, but violates
energy. Overall (``global'') energy conservation is then restored through the
rescaling procedure, which not only changes (reduces) the 3--momenta of all
particles in the final state but also gives rise to a flow of energy between
different parts of the event.  This class of algorithms is similar to that
employed for the fragmentation of heavy quarks in the ISAJET event simulation
package \cite{8}. It can be combined with either definition (\ref{e5}) or
(\ref{e6}) of the fragmentation variable; we will call the two resulting
algorithms I$_E$ and I$_k$, respectively.

Note that the sum of the square of the 4--momenta of a stop ``meson'' and its
associated fragmentation jet after rescaling exceeds $m^2_{\tilde{t}_1}$ in
the prescriptions $I_E$ and $I_k$. In other words, one implicitly assumes that
the stop squarks were produced slightly off--shell when implementing
fragmentation.\footnote{The matrix elements for stop production are
nevertheless computed assuming \st\ and $\tilde{t}_1^*$ to be on--shell, so as
to reproduce the correct {\em inclusive} \tp\ production cross section, in
accordance with QCD factorization theorems. The same apparent contradiction is
built into all QCD fragmentation or parton shower programs we know of.} Our
second algorithm makes this explicit, by directly generating a distribution in
the squared (off--shell) mass $m^2$ of the produced stop squarks:
\be \label{e7}
m^2(x) = m^2_{\tilde{t}_1} + 2 E^2_{\tilde{t}_1} (1-x) \left( 1 -
\sqrt{ 1 - \frac {m^2_{\tilde{t}_1}} {E^2_{\tilde{t}_1}} } \right) \;\;,
\ee
where $x$ is distributed according to the fragmentation function
$D_{\tilde t}(x)$ of eq.~(\ref{e1}), and $E_{\tilde{t}_1}$ is the
energy of {\em on--shell} stops, as generated during the calculation
of the stop production matrix element. The form (\ref{e7}) has been
obtained by writing the stop 4--momenta prior to and after fragmentation
as $(E_{\tilde{t}_1}, \vec{k})$ and $(x E_{\tilde{t}_1}, \vec{k}')$,
respectively, with $\vec{k}'$ being parallel to $\vec{k}$. One then
requires that the difference of these two 4--momenta is light--like
(for a massless fragmentation jet), and that the stop meson has mass
\mt. This allows to express $m^2$ as a function of $E_{\tilde{t}_1}$ and
$\vec{k}$. In the last step one replaces these quantities by the
energy and 3--momentum of on--shell \st; this can be regarded as a
one--step iteration. Had we instead defined the stop ``meson'' 4--momentum
as $(E', x \vec{k})$, i.e. tried to implement definition (\ref{e6}) in this
scheme via a one--step iteration, the second term on the r.h.s. of
eq.~(\ref{e7}) would have to be multiplied by an additional factor
$\sqrt{1 - m^2_{\tilde{t}_1} / E^2_{\tilde{t}_1} }$. However, we will
stick to the ansatz (\ref{e7}) for scheme II here, since the difference
between the definitions (\ref{e5}) and (\ref{e6}) of the scaling variable
will be explored by comparing the results of schemes I$_E$ and I$_k$.

Once the masses of \st\ and $\tilde{t}_1^*$ have been generated according to
eq.~(\ref{e7}), the energies and 3--momenta of all particles in the final
state have to be recomputed, using new limits of phase space as appropriate
for the increased stop masses. In so doing, care has to be taken to generate
kinematic quantities that ``correspond'' to those used (prior to
fragmentation) to compute the squared matrix element for stop production,
which enters the event weight. This can be accomplished as follows. Assume the
phase space is parameterized by a set of kinematic quantities $q_i, \ i = 1,
\dots, N$. The kinematic limits of $q_i$ are determined by \rs, by the masses
of \st\ and $\tilde{t}_1^*$, and (for $i \ne 1$) by the values of $q_1, \dots,
q_{i-1}$. An event can then be characterized by a set of dimensionless numbers
$r_i \in [0,1]$, so that $q_i = q_{i,{\rm min}} + r_i (q_{i,{\rm max}} -
q_{i,{\rm min}})$; the $r_i$ can be viewed as the random numbers used by a
Monte Carlo integration of phase space. The correspondence between
``fragmented'' and ``unfragmented'' events can then be ensured by using the
same set $\{ r_i \}$ when regenerating all 4--momenta after fragmentation,
only modifying the kinematical limits to accommodate increased stop squark
masses for the ``fragmented'' events. Once the 4--momenta of the off--shell
squarks have been generated in this way, fragmentation itself can be treated
as the (collinear) decay of a particle with mass $m$ into a second particle
with mass \mt\ and a massless parton.

This second algorithm has both advantages and disadvantages. On the
positive side, there is no need for iterative rescaling of 3--momenta
after fragmentation; however, as described in the previous paragraph,
all 4--momenta do need to be recomputed, allowing for different
(virtual) \st\ and $\tilde{t}_1^*$ masses. Another potential advantage
is that, although the ansatz (\ref{e7}) has been designed to emulate a
conventional fragmentation function using definition (\ref{e5}) of the
scaling variable, it could easily be converted into a fully Lorentz
invariant fragmentation scheme, since the mass $m^2$ is a Lorentz
invariant. This would be more difficult to accomplish in both variants
of scheme I, which are based on non Lorentz covariant quantities
(energies and 3--momenta). On the other hand, at least as implemented
using eq.~(\ref{e7}), scheme II does not have a smooth limit as $\mt
\rightarrow 0$. Eq.~(\ref{e7}) then always gives $m^2=0$, independent
of $x$, leaving the 4--momenta of the fragmentation products undefined. In
contrast, in the massless limit schemes I$_E$ and I$_k$ both approach the same
scheme (corresponding to independent fragmentation, again as in ISAJET
\cite{8}), where no rescaling of 3--momenta is required.

Rather than working with variable (virtual) stop squark masses and fixed
stop ``meson'' mass, one might also consider keeping the stop squark
masses fixed at \mt, but varying the stop meson masses. This would also
allow to implement energy--momentum conservation without any rescaling.
However, there are several problems with such a scheme. Energy--momentum
conservation would force the ``mesons'' to be lighter than the squarks,
whereas flavored $D$ and $B$ mesons are heavier than $c$ and $b$ quarks.
Also, a continuous stop meson energy distribution would correspond to
a continuous stop meson mass spectrum as well. For these reasons we 
will not explore this alternative any further.

In Figs.~1a,b we compare results for our three fragmentation schemes. We have
chosen a parameter set of current interest, $\rs=183$ GeV and $\mt=80$
GeV. The solid, long dashed and short dashed curves refer to schemes I$_k$,
I$_E$ and II, respectively. We have taken \st\ to be a pure $SU(2)$ doublet
($\theta_{\tilde t} = 0$ in the notation of refs.~\cite{7,9}), but this only
affects the overall normalization of the curves. We have also included ``soft
and virtual'' QCD corrections \cite{9}, defined through an infrared cut--off
for emitted gluons of 1 GeV. This ensures that we are only studying events
with $2 \rightarrow 2$ kinematics; events with additional hard gluon will be
discussed in Sec.~3.

\begin{figure}[htb]
\vspace*{-4.5cm}
\centerline{\epsfig{file=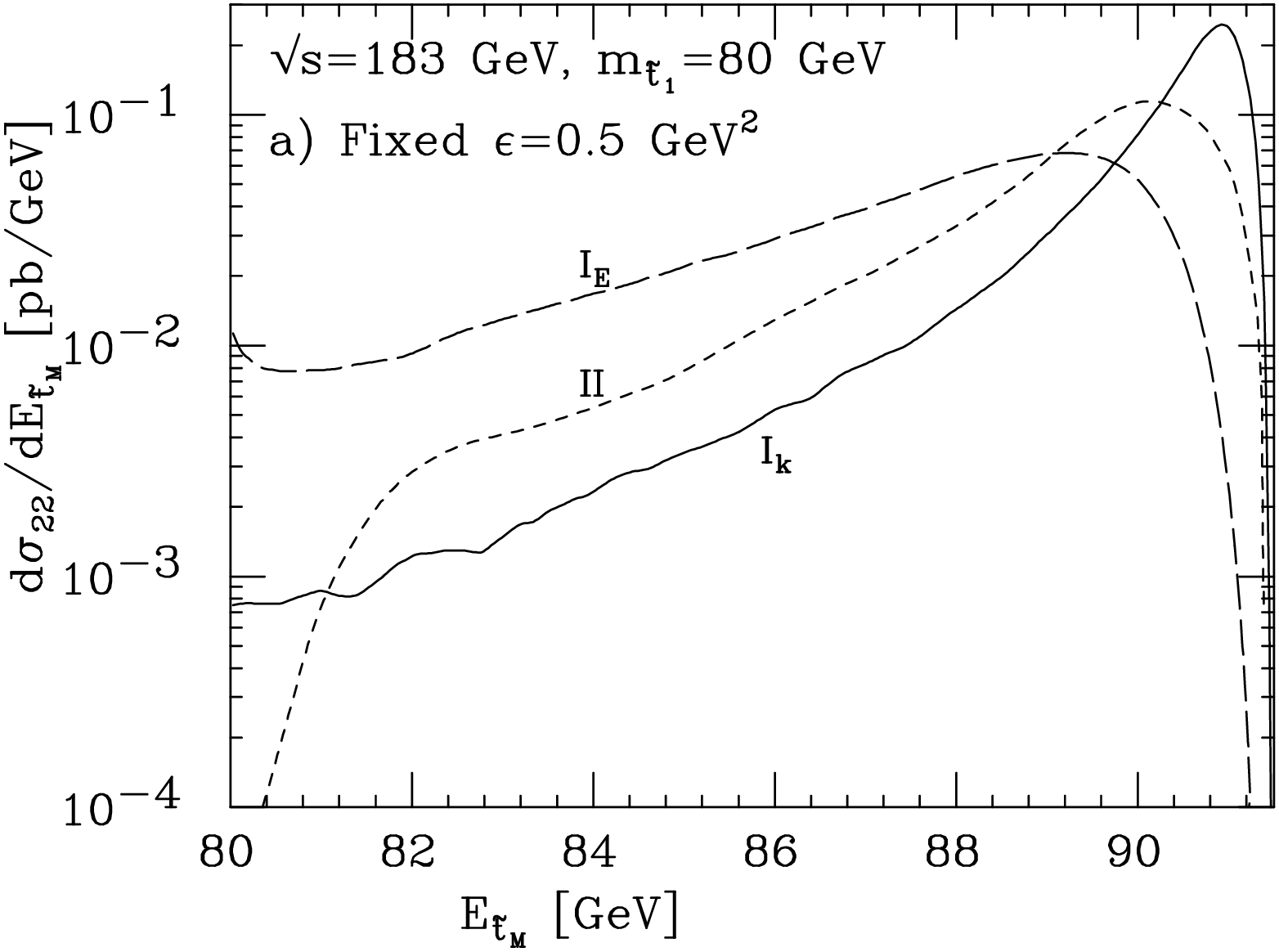,height=13.3cm}}
\vspace*{-4cm}
\centerline{\epsfig{file=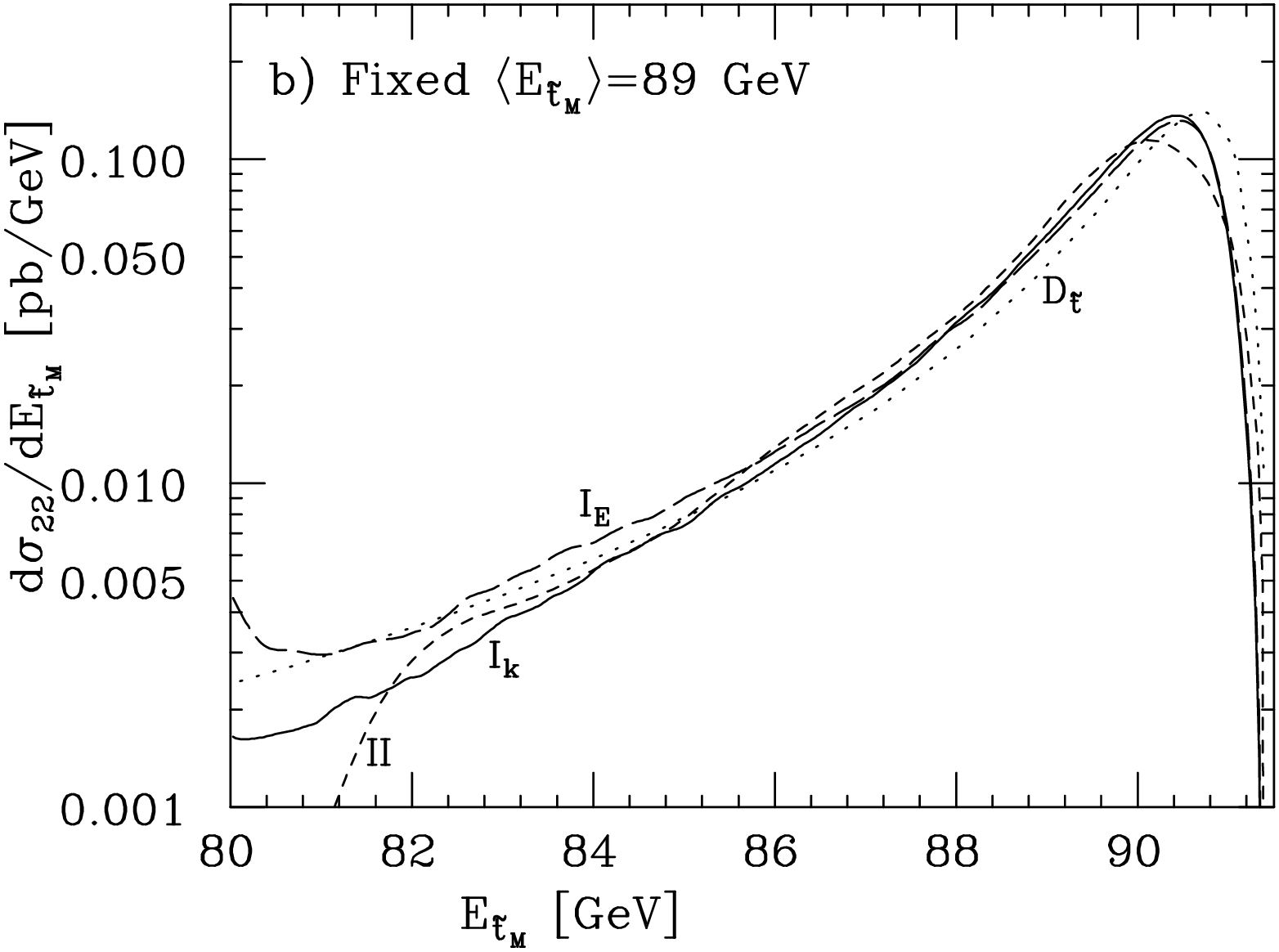,height=13.3cm}}

\caption
{The stop ``meson'' spectrum as predicted by the three fragmentation schemes
described in the text. All three schemes employ the Peterson {\it et al.} 
fragmentation function. In a) the same value of the parameter $\epsilon$ has
been used in all cases, while in b) the value of $\epsilon$ has been adjusted
so as to reproduce the prediction $\langle E_{\tilde{t}_M} \rangle \simeq 89$
GeV of the original Peterson {\it et al.}  fragmentation function with
$\epsilon=0.5$ GeV$^2$, using the definition (\ref{e5}) for the scaling
variable; this gives $\epsilon = 0.1, 2.0$ and $0.5$ GeV$^2$ for schemes
I$_E$, I$_k$ and II, respectively. The shape of the Peterson {\it et
al.} fragmentation function is shown by the dotted curve.}

\end{figure}

In Fig.~1a we compare the stop ``meson'' spectra produced by our three
schemes. We have taken the same value of $\epsilon$ in all three cases,
but the resulting spectra clearly differ substantially. The difference
between schemes I$_E$ and I$_k$ is fairly easy to understand. Since the
produced stop squarks are far from being ultra--relativistic ($\beta
\simeq 0.5$), reducing their 3--momenta by some factor $x$ corresponds
to a much smaller relative reduction of their total energies, which
receive a large (constant) contribution from their mass. As a result,
for fixed $\epsilon$ scheme I$_k$ leads to a much harder stop ``meson''
spectrum than scheme I$_E$ does. Except for the steep drop towards the
lower kinematical endpoint of the distribution, scheme II falls in between
the two variants of scheme I here. The large differences displayed in
this figure clearly show that care must be taken when comparing the
values of the free parameters of heavy (s)quark fragmentation functions
obtained from different event simulation programs, with possibly different
implementations of the fragmentation process.

On the other hand, Fig.~1b shows that at least for the simple case of
exclusive \tp\ production ($2 \rightarrow 2$ kinematics), the differences
between fragmentation schemes can largely be absorbed into a change of the
value of the parameter $\epsilon$ of eq.~(\ref{e2}). Here we have chosen
$\epsilon$ in each scheme such that the average stop meson energy coincides
with the prediction of the original Peterson {\it et al.} function (\ref{e1});
using $\epsilon = 0.5$ GeV$^2$ and the definition (\ref{e5}) of the scaling
variable, it predicts $\langle E_{\tilde{t}_M} \rangle = 89.1$ GeV. Almost no
change of $\epsilon$ is required in scheme II. On the other hand, in scheme
I$_E$, $\epsilon$ has to be reduced to $0.1$ GeV$^2$, while in scheme I$_k$ it
has to be increased to $2.0$ GeV$^2$, in order to reproduce the same $\langle
E_{\tilde{t}_M} \rangle $. These two values of $\epsilon$ differ by a factor
of 20!

After this adaptation of $\epsilon$, the three schemes give quite similar
results for the stop ``meson'' spectrum, except near its lower kinematical
boundary. However, recall that these schemes ensure energy--momentum
conservation only globally. This means that the fragmentation process will
change the 4--momenta of partons that seemingly have nothing to do with
it. This is actually not as unreasonable as it may sound, since color
neutralization in any case connects different partons during hadronization.
Furthermore, notice that none of the three schemes reproduces the original
Peterson function, shown as the dotted curve, near the upper endpoint of the
spectrum. The reason is that, due to the requirements of overall
energy--momentum conservation, our schemes allow $E_{\tilde{t}_M} \simeq
\rs/2$ only if {\em both} fragmentation variables are close to 1, which is
obviously much less likely than having only one of them close to unity.  We
will see in the next Section that even after adapting the value of $\epsilon$
so as to produce the same $\langle E_{\tilde{t}_M} \rangle$, the three schemes
lead to somewhat more pronounced differences in events with additional hard
gluon.


\section*{3) Hard gluon emission}

\setcounter{footnote}{0}
In this Section we discuss the production of a \tp\ pair in association with a
hard gluon, $\epem \rightarrow \tp g$. The gluon can be emitted from either of
the two stop lines; in addition, there are diagrams with a $\gamma \tp g$ or
$Z \tp g$ vertex. This process has been treated previously in
refs.~\cite{9,abdel,ken}, using parameterizations of phase space that simplify
analytical calculations. In contrast, our treatment is geared towards easy
implementation in an event generator program. The spin--averaged squared
matrix element is
\clearpage
\beq \label{e8}
&\left| \overline{\cal M}\left( e^+(p_1)~ e^-(p_2) \rightarrow \st(k_1)~
\tilde{t}_1^*(k_2)~ g(q) \right) \right|^2 
\nonumber \\
&= 4 \frac{g_s^2 e^4} {s^2} \left( |A|^2 + |B|^2 \right) \left\{ 
4s - 2 m^2_{\tilde{t}_1} \left[ \frac { 4k_1 \cdot p_1~ k_1 \cdot p_2 - s~
m^2_{\tilde{t}_1} } { ( k_2 \cdot q )^2 } +
\frac { 4k_2 \cdot p_1~ k_2 \cdot p_2 - s~ m^2_{\tilde{t}_1} } 
{ ( k_1 \cdot q )^2 } \right]
\right. \nonumber \\ & \left. \hspace*{40mm}
+  \frac { k_1 \cdot q + k_2 \cdot q + k_1 \cdot k_2 } 
{ k_1 \cdot q k_2 \cdot q }
\left[ s~ k_1 \cdot k_2 - 2 \left( p_1 \cdot k_1~ p_2 \cdot k_2
+ p_2 \cdot k_1~ p_1 \cdot k_2 \right) \right] \right\}
\nonumber \\
& \equiv 4 \frac {g_s^2 e^4} {s^2} \left( |A|^2 + |B|^2 \right)
\left| \widetilde{\cal M} \right|^2 \;\;,
\eeq
where $g_s$ is the strong coupling constant, which we take at scale
$\mu = \rs$, and $e$ is the QED coupling. The coefficients $A$ and $B$
describe the strength of the effective vector and axial vector couplings,
being given by
\ben \label{e9} \beq
A &= q_{\tilde t} - \frac { (1 - 4 \sin^2 \theta_W) c_{\tilde t} }
{8 \sin^2 \theta_W \cos^2 \theta_W} \cdot 
\frac {s} {s - M_Z^2 + i M_Z \Gamma_Z} \;\;;
\label{e9a} \\
B &= \frac { c_{\tilde t} } {8 \sin^2 \theta_W \cos^2 \theta_W} 
\cdot \frac {s} {s - M_Z^2 + i M_Z \Gamma_Z} \;\;,
\label{e9b}
\eeq \een
where
\be \label{e10}
c_{\tilde t} = 2 q_{\tilde t} \sin^2 \theta_W - 2 \cos^2 \theta_{\tilde t}
I_{3_{\tilde t}}
\ee
describes the $Z \tp$ coupling \cite{7} with $\theta_{\tilde t}$ being
the stop mixing angle (the angle between \st\ and $\tilde{t}_L$);
$q_{\tilde t}$ ($= +2/3$) and $I_{3_{\tilde t}}$ ($=+1/2$) are the
\st\ charge and $\tilde{t}_L$ weak isospin, respectively.

The cross section can now be expressed in terms of the reduced squared matrix
element $\left| \widetilde{\cal M} \right|^2$ introduced in eq.~(\ref{e8}):
\be \label{e11}
\sigma(\epem \rightarrow \tp g) = \frac {\alpha_s \alpha_{em}^2 }
{\pi s^3} \left( |A|^2 + |B|^2 \right) \int dE_g~ d \cos \theta_g~ 
d E_{\tilde t}~ d \phi_{\tilde t}~ \left| \widetilde{\cal M} \right|^2 \;\;.
\ee
Here, $E_g$ and $E_{\tilde t}$ are the energies of the gluon and of \st\
in the \epem\ cms frame, $\theta_g$ is the angle between the
gluon and the direction of the positron beam, and $\phi_{\tilde t}$
determines the direction of the \st\ 3--momentum, as follows. We choose
the frame where the gluon's 3--momentum is given by
\be \label{e12}
\vec{q} = E_g ( 0, \sin \theta_g, \cos \theta_g) \;\;,
\ee
{\it i.e.} the integration over the polar angle of the gluon has been
performed explicitly in eq.~(\ref{e11}), yielding a factor of $2 \pi$. The
\st\ 3--momentum is then given by
\be \label{e13}
\vec{k}_1 = \sqrt{ E_{\tilde t}^2 - m_1^2 } ( \sin \phi_{\tilde t}
\sin \alpha, \cos \theta_g \cos \phi_{\tilde t} \sin \alpha +
\sin \theta_g \cos \alpha,
\cos \theta_g \cos \alpha - \sin \theta_g \cos \phi_{\tilde t} \sin \alpha ),
\ee
where
\be \label{e14}
\cos \alpha = \frac { s - 2 \sqrt{s} (E_g + E_{\tilde t}) + 2 
E_g E_{\tilde t} + m_1^2 - m_2^2 } {2 E_g \sqrt{ E^2_{\tilde t} - m_1^2 } }
\ee
is the angle between \st\ and the gluon, which is fixed for given $E_g$
and $E_{\tilde t}$. Note that we have allowed for different \st\ and
$\tilde{t}_1^*$ masses $m_1$ and $m_2$ in eq.~(\ref{e14}), as required for
the implementation of our fragmentation scheme II discussed in the
previous section. However, for the purpose of computing event weights
$m_1 = m_2 = \mt$ should be used. The $\tilde{t}_1^*$ 3--momentum can
now trivially be computed from eqs.~(\ref{e12})--(\ref{e14}) as
$\vec{k}_2 = -\vec{k_1} - \vec{q}$.

Finally, the limits of phase space integration, again allowing for
different \st\ and $\tilde{t}_1^*$ masses, are:
\ben \label{e15} \beq
E_g & \in [ E_g^{\rm min},  \frac {s - (m_1 + m_2)^2 } {2 \sqrt{s} } ] \;\;;
\label{e15a} \\
\cos \theta_g & \in [-1, 1] \;\;;
\label{e15b} \\
\phi_{\tilde t} & \in [0, 2 \pi] \;\;;
\label{e15c} \\
E_{\tilde t} & \in [E_-,  E_+] \;\;,
\label{e15d}
\eeq \een
where
\be \label{e16}
E_\pm = \frac {\sqrt{s} - E_g} {2} \left[ 1 + 
\frac { m_1^2 - m_2^2 } { \sqrt{s} ( \sqrt{s} - 2 E_g ) } \right]
\pm \frac {E_g}{2} \sqrt{ 
\left[ 1 + \frac { m_1^2 - m_2^2 } {\sqrt{s} ( \sqrt{s} - 2 E_g ) }
\right]^2 - \frac {4 m_1^2} {\sqrt{s} ( \sqrt{s} - 2 E_g ) } } \;\;.
\ee
Note that the squared matrix element (\ref{e8}) diverges linearly as $E_g
\rightarrow 0$. This requires the introduction of an infrared cut--off
$E_g^{\rm min}$ in eq.~(\ref{e15a}). The same cut--off also appears
(logarithmically) in the expression for the ``soft $+$ virtual'' QCD
corrections. Physical quantities do not depend on $E_g^{\rm min}$, if it is
chosen sufficiently small so that events containing a gluon with energy $E_g =
E_g^{\rm min}$ still effectively possess $2 \rightarrow 2$ kinematics. We have
checked numerically that eqs.~(\ref{e8})--(\ref{e16}) reproduce the ``hard''
real QCD corrections to the total \tp\ production cross section as given in
ref.~\cite{9} (and contained implicitly in \cite{7}). Of course, stop
fragmentation, as implemented using either of our schemes, does not change the
total cross section before cuts.

\begin{figure}[htb]
\vspace*{-3.5cm}
\centerline{\epsfig{file=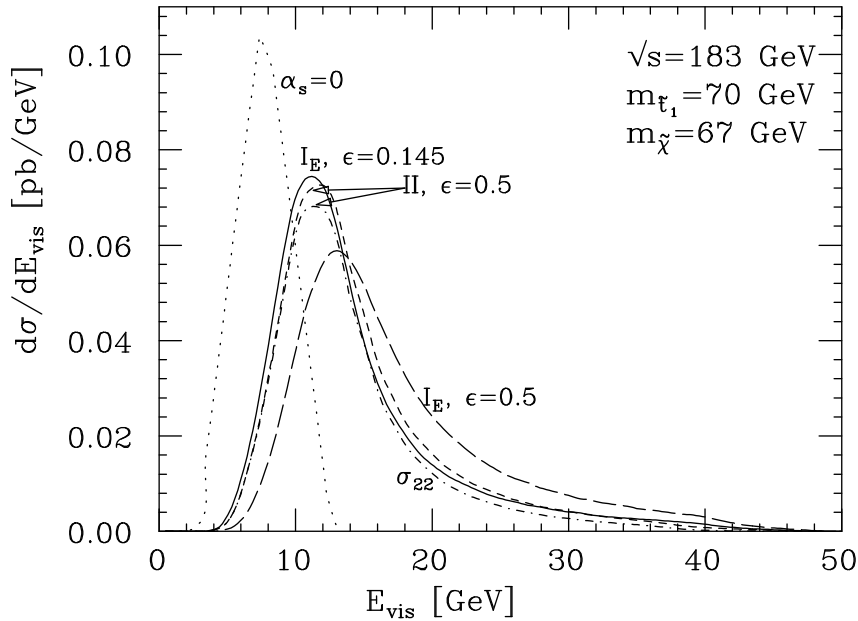,height=13.3cm}}

\caption
{The total visible energy in $\epem \rightarrow \tp (g)$ events at
LEP--183. The dotted curve has been obtained by switching off both
perturbative and non--perturbative strong interaction effects, while the
dot--dashed curve includes non--perturbative \st\ fragmentation but no hard
gluon radiation (defined with an infrared cut--off $E_g^{\rm min} = 1$
GeV). The other curves include both perturbative and non--perturbative
effects. The labels refer to the fragmentation scheme used, and give the value
of the fragmentation parameter $\epsilon$ of eq.~(\ref{e2}) in units of
GeV$^2$.}

\end{figure}

In the subsequent analysis we have taken into account the \st\ decay, which we
have assumed to be always into a massless quark\footnote{In models with exact
universality of the squark masses at the scale $M_X$ of Grand Unification one
expects \st\ to dominantly decay into a charm quark \cite{5}; in this case the
visible energy could be further degraded by semileptonic decays of the charm
(anti)quark. However, there is no compelling reason to assume exact (rather
than only approximate) squark degeneracy at $M_X$.  Moreover,
\st\ decays are independent of the strong interaction effects on \tp\
pair production, which are the main focus of our analysis.} plus an invisible
LSP \tchi.  A first illustration of the impact of both perturbative and
non--perturbative strong interaction effects on \tp\ production is shown in
Fig.~2, where we plot the total visible energy for $\rs=183$ GeV, $\mt=70$
GeV, and $\mchi=67$ GeV. We have chosen a scenario with very small \st--LSP
mass splitting, since in this case strong interaction effects are most
important. This is illustrated by the difference between the dotted curve,
where all such effects have been switched off, and the other curves.  In the
absence of strong interaction effects, and before any cuts have been applied,
the visible energy spectrum has a triangular shape, centered at $E_q^* \cdot
(\rs/\mt) \simeq 7.7$ GeV, where $E_q^* = (m^2_{\tilde{t}_1} - m^2_{\tilde
\chi})/(2 \mt)$ is the quark energy in the \st\ rest frame. We see that strong
interaction effects are expected to nearly double the average visible energy
for our choice of sparticle masses.

However, there are significant uncertainties in this prediction, due to our
lack of understanding of non--perturbative \st\ fragmentation. Different
fragmentation schemes using the same fragmentation function can give quite
different results, as can be seen by comparing the long and short dashed
curves, which are valid for schemes I$_E$ and II, respectively, with $\epsilon
= 0.5$ GeV$^2$. Such a difference is not surprising, given the results of
Fig.~1. However, even after adjusting the value of $\epsilon$ used in scheme
I$_E$ so as to reproduce the average stop ``meson'' energy predicted by scheme
II, which is very close to the value predicted directly from the Peterson {\it
et al.} fragmentation function, some discrepancies remain. Indeed, the
difference between the predictions of scheme I$_E$ with adjusted $\epsilon$
parameter (solid curve) and scheme II is comparable in size to the effect of
hard gluon radiation, which can be seen by comparing the short dashed and
dot--dashed curves; the former shows the full cross section, while the latter
is for the $2 \rightarrow 2$ contribution only.

One might be tempted to conclude from Fig.~2 that non--perturbative stop
fragmentation by itself can release a sufficient amount of visible energy to
give a viable signal for \tp\ production even for small
\st--LSP mass splittings. However, one should keep in mind that in the absence 
of hard gluon radiation, the fragmentation products will always be almost
back--to--back in the transverse plane, and the \st\ decay products are not
energetic enough to change this picture. Unfortunately there are large
backgrounds to such configurations, even if one requires a substantial
transverse momentum imbalance. The main source of backgrounds are two photon
processes. They can lead to large (apparent) missing $p_T$ either through
semileptonic decays of charm quarks, or through mismeasurement since the
energies of the jets are quite low, so that calorimetry becomes rather
imprecise. Recent experimental searches \cite{3,4} therefore discard such
back--to--back configurations. This will enhance the importance of signal
events containing an additional hard gluon.

\begin{figure}[htb]

\vspace*{-3cm}

\centerline{\epsfig{file=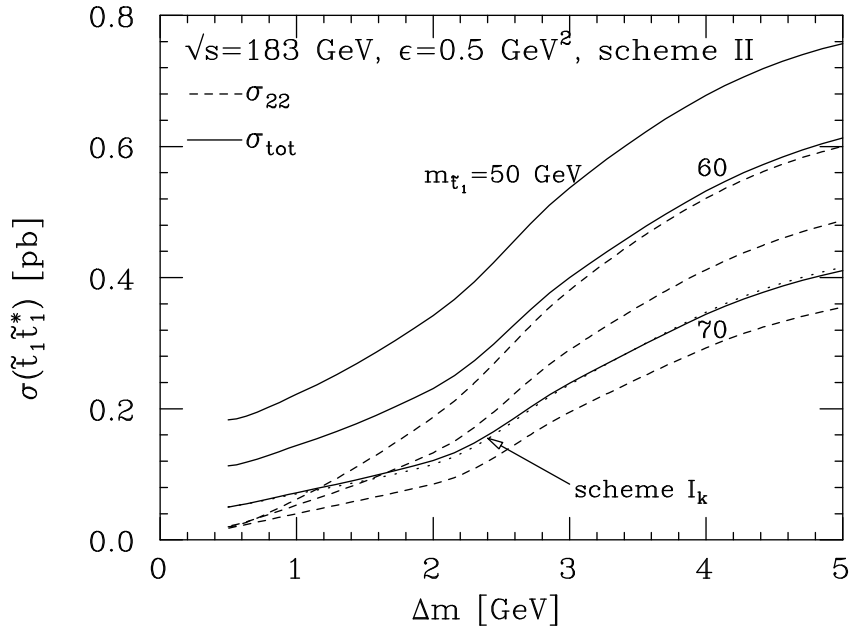,height=13.3cm}}

\caption
{The total inclusive \tp\ cross section at LEP--183 after the cuts (\ref{e17})
have been applied. The solid curves show the full result, while the dashed
curves have been computed by switching off hard gluon radiation. For $\mt=50$
and 60 GeV, the infrared cut--off has been chosen such that the total cross
section before cuts of the $2 \rightarrow 2$ contribution reproduces the
leading order, ${\cal O}(\alpha_s^0)$, prediction; this gives $E_g^{\rm
min}=5.4 \ (1.8)$ GeV for $\mt = 50 \ (60)$ GeV. For $\mt=70$ GeV this would
give a very small value of $E_g^{\rm min}$, so we have chosen 1.0 GeV
instead. Recall that the total result (solid curves) does not depend on
$E_g^{\rm min}$. All curves are for fragmentation scheme II, except the
dotted one, which is for scheme I$_k$.}

\end{figure}

This is illustrated in Fig.~3, where we have attempted to incorporate the
most important cuts applied by the OPAL collaboration \cite{4} into our
parton level event generator. We show the total inclusive \tp\ cross
section at $\rs=183$ GeV after the following cuts have been applied:
\ben \label{e17} \beq
M_{\rm vis} &> 4.0 \ {\rm GeV} \;\;;
\label{e17a} \\
p_{T,{\rm miss}} &> 4.5 \ {\rm GeV} \;\;;
\label{e17b} \\
|\cos \theta_{\rm miss}| & < 0.9 \;\;;
\label{e17c} \\
\phi_{\rm acop} &> 20^\circ \;\;.
\label{e17d}
\eeq \een
In order to model the geometric acceptance of a typical LEP detector, we have
discarded all partons with azimuthal angle $|\cos \theta| > 0.97$ before
computing the total visible mass $M_{\rm vis}$ and missing transverse momentum
$p_{T,{\rm miss}}$. The cut (\ref{e17c}) is on the angle of the missing
momentum vector with the beam pipe. OPAL also requires that the $k_T$
clustering algorithm \cite{10} with fixed jet resolution parameter $y_{\rm
cut}$ produce exactly two jets, and applies cuts on the invariant masses of
these jets. However, it is difficult to model such cuts accurately with a
parton level event generator. Furthermore, insisting on exactly two jets with
given $y_{\rm cut}$ would tend to remove events containing a hard gluon in
addition to the \tp\ pair. Therefore, the full set of cuts applied by the OPAL
collaboration is not efficient for scenarios with very small $\st-\tchi$ mass
splitting. We have instead used the $k_T$ clustering algorithm to force the
event into a two--jet structure, except for events with less than two partons
in the acceptance region of the detector, which we discarded -- that is, we
kept merging jets until exactly two jets remain. We then applied the cut
(\ref{e17d}) on the angle between these two jets in the transverse plane.

The results of Fig.~3 indicate that the emission of a hard gluon should
allow to detect or exclude \tp\ pair production even for arbitrarily
small $\st-\tchi$ mass splitting. The published OPAL limit \cite{4} on
\mt\ corresponds to a total cross section of 0.13 pb, after the
cuts (\ref{e17}) have been applied. This indicates that existing data might
allow to exclude, or find evidence for, a \st\ nearly degenerate with the LSP
up to $\mt \simeq 60$ GeV for mixing angle $\theta_{\tilde t} =0$. Of course,
a complete experimental analysis would be needed to really establish such a
bound. We emphasize that an accurate description of hard gluon radiation would
be crucial for such an analysis. Leading--log parton shower models, as used in
refs.~\cite{3,4}, cannot be expected to model this process accurately. The
importance of this $2 \rightarrow 3$ contribution is illustrated by the dashed
curves, where this contribution has been turned off; the cross section after
cuts then goes to zero in the limit of vanishing $\st-\tchi$ mass difference.

Most results shown in Fig.~3 have been computed using scheme II with $\epsilon
= 0.5$ GeV$^2$. However, the cuts (\ref{e17}) greatly reduce the sensitivity
to the details of the fragmentation process. This is illustrated by the dotted
curve, which shows the inclusive \tp\ cross section after cuts for $\mt = 70$
GeV, as predicted using scheme I$_k$ with the {\em same} value of
$\epsilon$. We saw in Fig.~1 that this corresponds to a significantly harder
stop ``meson'' spectrum, {\it i.e.} much softer fragmentation
jets. Nevertheless the predicted cross section is almost the same as that for
scheme II. This is due to a somewhat fortuitous cancellation: models with
harder fragmentation jets predict a significantly larger cross sections after
the cuts (\ref{e17}a,b) have been applied; however, the resulting event then
also tends to be more back--to--back, so that a relatively larger number of
events is removed by the cut (\ref{e17d}). This weak sensitivity to the
details of the fragmentation process has also been noticed in the experimental
searches \cite{3,4}, where it was studied by varying the parameter $\epsilon$
of the Peterson {\it et al.}{ fragmentation function. This result is very
welcome, since it almost removes one (potentially large) source of uncertainty
of the measured (limit on the) total \tp\ cross section.

As already mentioned in the Introduction, events with an additional
hard gluon might also help to determine the color charge of
\st.\footnote{The ``3--jet'' contribution to the total inclusive cross
section for the production of a pair of colored scalars has been
calculated in \cite{ken}; however, that analysis did not treat
fragmentation and decay of the scalars, nor were any cuts applied to
reject the background.} Given the present experimental limits, a
definitive analysis of this kind may only be possible at a high
energy, high luminosity linear \epem\ collider (LC)\footnote{Of
course, the discovery of \st\ at a hadron collider would immediately
prove that it carries a nonvanishing color charge. However, in order
to confirm that it belongs to an $SU(3)_C$ triplet one would have to
measure both \mt\ and the \tp\ production cross section, which may not
be easy.}. Various designs of such colliders are now being studied,
with quoted luminosities ranging up to 500 fb$^{-1}$/yr \cite{11}.

Assuming that \st\ always undergoes 2--body decay into a light quark and an
invisible LSP \tchi, the most direct way to determine its color charge is
probably through the study of events with 3 or more jets.  Notwithstanding,
non--perturbative \st\ fragmentation can also produce jets. Fortunately these
fragmentation jets are expected to be quite soft; perturbative contributions
should therefore dominate for sufficiently large jet energies.

This is borne out by Fig.~4, which shows results of an analysis for $\rs =
500$ GeV, $\mt = 150$ GeV, and $\mchi = 100$ GeV. Allowing for the (probably)
larger beam holes of future LC detectors as compared to present LEP detectors,
we have discarded all partons with $|\cos \theta| > 0.9$.  We then used the
$k_T$ clustering algorithm \cite{10} to force the event into a 3--jet
structure. Following ref.~\cite{12}, we demanded that the hardest two jets
have energies $E_{{\rm jet}_{1,2}} > 15$ GeV, and that the total missing $p_T$
exceeds 35 GeV; this removes most SM backgrounds.  Fig.~4 shows the energy
spectrum of the third (softest) jet after these cuts.

\begin{figure}[htb]

\vspace*{-3cm}

\centerline{\epsfig{file=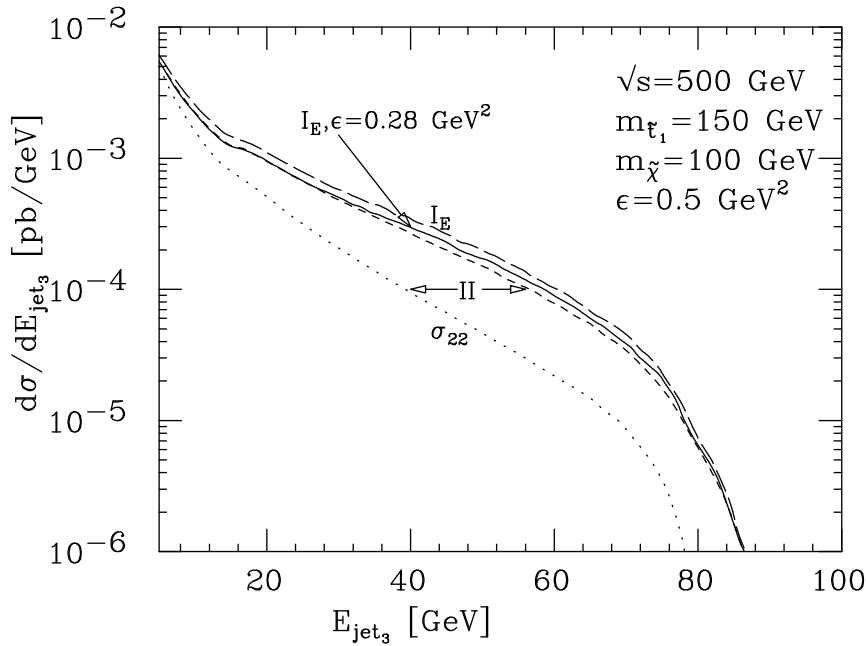,height=13.3cm}}

\caption
{The spectrum of the softest jet produced in \tp\ production at a future
linear collider after the cuts described in the text, where the events have
been forced into a 3--jet topology using the $k_T$ clustering algorithm. The
small shoulder at $E_{{\rm jet}_3} = 15$ GeV is caused by the cut on the
energies of the two hardest jets. The dotted curve shows the result in the
absence of hard gluon radiation, as predicted using fragmentation scheme II,
while the other three curves show the total inclusive cross section. They
differ in the fragmentation scheme and/or in the value of the fragmentation
parameter $\epsilon$, as indicated.}

\end{figure}

The dotted curve shows the contribution of events without hard gluon (defined
through an IR cut--off $E_g^{\rm min}=5$ GeV), while the other three curves
include contributions from both $2 \rightarrow 2$ and $2 \rightarrow 3$
processes. We see that for $E_{{\rm jet}_3}>15$ GeV, the cross section is
dominated by perturbative contributions. However, a significant dependence on
the details of non--perturbative \st\ fragmentation remains even in this
region. We saw in Fig.~1 that for a given value of $\epsilon$, scheme I$_E$
predicts a softer stop ``meson'' spectrum, {\it i.e.} harder fragmentation
jets, than scheme II does. It is therefore not surprising that scheme I$_E$
predicts the third jet to be somewhat more energetic. However, some difference
between the predictions from these two schemes remains even if the value of
$\epsilon$ is adjusted to reproduce the same average stop ``meson''
energy. After this adjustments the two schemes make very similar predictions
for $E_{{\rm jet}_3} \leq 30$ GeV, where most events still have an effective
$2 \rightarrow 2$ kinematics. Nevertheless, our different methods to ensure
overall energy--momentum conservation still give somewhat different
predictions for $E_{{\rm jet}_3} > 30$ GeV, where most events contain a quite
energetic gluon.

These differences are partly due to the fact that we force all events
into a 3--jet structure, so that the gluon is often merged with (some
of) the fragmentation products. This source of dependence on the
fragmentation scheme can probably be reduced if a different jet
definition is used.  On the other hand, the two schemes also predict a
somewhat different gluon energy spectrum even prior to any jet
merging. We therefore expect that some dependence on the fragmentation
scheme will remain for all jet definitions. Of course, for the
parameters chosen in Fig.~4, the large number of \tp\ pair events
($\sim 65,000$ signal events after cuts for an integrated luminosity
of 500 fb$^{-1}$) should also allow to study various aspects of
non--perturbative \st\ fragmentation in detail. Moreover, since the
differences between the predictions of our fragmentation schemes are
far smaller than the overall contribution of events containing hard
gluons, even our current level of understanding of non--perturbative
\st\ fragmentation would suffice to ascertain with a high degree of
confidence that \st\ is indeed a color triplet.


\section*{4) Summary and Conclusions}

We have examined the effects of both perturbative and
non--perturbative strong interactions on \tp\ production at current
and future \epem\ colliders. The latter manifests itself in \st\
fragmentation into a stop ``meson'', while the former gives rise to
events containing a hard gluon in addition to the stop squarks.

Non--perturbative \st\ fragmentation has been discussed in some detail
in Sec.~2, with particular emphasis on the conceptual ambiguities of
the fragmentation of heavy partons. These ambiguities are largest for
beam energies somewhat, but not far, above the pair production
threshold, {\it i.e.} precisely in the region where squark production
can best be studied at \epem\ colliders. They include not only the
definition of the fragmentation (scaling) variable, but also the
algorithm used to ensure overall energy--momentum conservation. The
scaling variable can be defined either through the energy or through
the absolute value of the 3--momentum (or through some combination
thereof). These two quantities coincide for massless partons, and also
for massive partons in the infinite momentum frame where QCD
factorization theorems are often formulated. However, they can differ
greatly for massive partons in the laboratory frame, where event
generators are needed to describe the data. We also introduced two
different schemes to guarantee overall energy and momentum
conservation after fragmentation. The first scheme is based on the
re--scaling of all 3--momenta, and is quite similar to algorithms
employed in multi--purpose event generators like ISAJET. In the second
scheme, which bears some resemblance to algorithms often used to model
perturbative parton showering, one explicitly generates a distribution
in the virtuality of the produced squarks. We saw that these schemes
can lead to quite different results when applied naively to the same
fragmentation function.  However, at least in the simple case of
exclusive \tp\ production, {\it i.e.}  in the absence of hard gluon
radiation, most of these differences can be absorbed in a redefinition
of the parameter(s) of the fragmentation function.

Events containing hard gluons have been studied in Sec.~3. After
giving explicit expressions for the corresponding fully differential
cross section, we showed that such events should allow to extend the
sensitivity of \epem\ collider experiments to scenarios with
arbitrarily small mass splitting between \st\ and the lightest
superparticle \tchi, which we assume to be invisible. The gluon by
itself can produce a sufficient amount of visible energy, and of
missing transverse momentum and acoplanarity, to discriminate between
signal and background even if \st\ decays release almost no visible
energy. We found that the cuts used to suppress Standard Model
backgrounds also reduce the sensitivity of the predicted cross section
to the details of \st\ fragmentation to a negligible level. Hard gluon
radiation had previously been found \cite{13} to increase the
sensitivity of hadron colliders to \tp\ pair production in scenarios
with small stop--LSP mass splitting. However, that analysis treated
gluon radiation only in leading logarithmic approximation (through the
parton shower algorithm of ISAJET), rather than through exact matrix
elements. Moreover, scenarios with mass splitting below 20 GeV still
remained undetectable in hadronic collisions.

We have also indicated how an analysis of 3--jet events from \tp\ pair
production at a future linear collider operating at high energy and
high luminosity can be used to determine the color charge of \st. In
this case some sensitivity to \st\ fragmentation remains after cuts.

Our numerical results include both real and virtual ordinary QCD
corrections. However, we did not include initial state radiation, nor
did we include corrections from loops including gluinos, {\it i.e.}
``SUSY QCD corrections'' \cite{abdel,14}. Both effects mostly change
the overall normalization of the \tp\ production cross section,
leaving the shapes of the distributions studied here almost
unchanged.\footnote{Events where a hard photon is radiated down one of
the beam pipes will have significantly reduced visible
energy. However, the cross section for such events is quite small,
unless the collider operates at energies far above the stop
threshold.} The total cross section also depends on the $\tilde{t}_L -
\tilde{t}_R$ mixing angle $\theta_{\tilde t}$, which we set to zero in
our numerical examples.

The numerical analyses presented here are by no means an exhaustive list
of the effects of strong interactions on stop production at \epem\
colliders. For example, strong interactions will play a role in 
measurements of the stop and LSP masses from \tp\ events. This is
true in particular in scenarios with small stop--LSP mass splitting,
where a large fraction of the total visible energy can come from
non--perturbative \st\ fragmentation or hard gluon radiation, as
illustrated in Fig.~2. Even if this mass splitting is quite large,
the extracted values of these masses could be shifted by several GeV
if the analysis is not designed such as to minimize these effects.

While the other very short--lived squarks are not expected to undergo
non--perturbative fragmentation, hard gluon radiation will occur with
equal relative frequency for all squark flavors. The cross section
given in Sec.~3 can easily be adapted to the production of other
squarks by changing the relevant electroweak couplings. Many SUSY
models predict first and second generation squarks to be closely
degenerate in mass; moreover, $SU(2)$ doublet squarks might undergo
quite complicated cascade decays even if they are lighter than the
gluino. One will then need a more sophisticated analysis to
experimentally determine their color charges. The effect of hard gluon
radiation on squark mass measurements in such a scenario \cite{12}
also needs to be investigated. On the other hand, since \st\ might
well be very long--lived compared to the time scale of strong
interactions in spite of being very heavy in relation to the QCD scale
$\Lambda$, it might even provide us with a unique tool to study
non--perturbative aspects of the hadronization process. Analyses of
squark pair production at \epem\ colliders can thus not only yield
valuable information on these (as yet hypothetical) new particles, but
might even improve our understanding of some basic aspects of strong
interactions.


\subsection*{Acknowledgements}
MD acknowledges financial support from FAPESP (Brazil). OJPE gratefully
acknowledges the hospitality of the Phenomenology Institute of the University
of Wisconsin. This work was supported in part by the University of Wisconsin
Research Committee with funds granted by the Wisconsin Alumni Research
Foundation, by the U.S.\ Department of Energy under grant DE-FG02-95ER40896,
by Conselho Nacional de Desenvolvimento Cient\'{\i}fico e Tecnol\'ogico
(CNPq), by Funda\c{c}\~ao de Amparo \`a Pesquisa do Estado de S\~ao Paulo
(FAPESP), and by Programa de Apoio a N\'ucleos de Excel\^encia (PRONEX).


\end{document}